\documentclass[aps,prb,showpacs,superscriptaddress]{revtex4}
\usepackage{amsmath}
\usepackage{amsfonts}
\usepackage{graphicx}

\begin{document}

\title{Microwave emission from a crystal of molecular magnets -- The role of
a resonant cavity}
\author{Mih\'{a}ly G. Benedict}
\affiliation{Department of Theoretical Physics, University of Szeged, Tisza Lajos k\"{o}r%
\'{u}t 84, H-6720 Szeged, Hungary}
\author{P\'{e}ter F\"{o}ldi}
\affiliation{Department of Theoretical Physics, University of Szeged, Tisza Lajos k\"{o}r%
\'{u}t 84, H-6720 Szeged, Hungary}
\author{F. M. Peeters}
\affiliation{Departement Fysica, Universiteit Antwerpen, Groenenborgerlaan 171, B-2020
Antwerpen, Belgium}

\begin{abstract}
We discuss the effects caused by a resonant cavity around a sample of a
magnetic molecular crystal (such as Mn${}_{12}$-Ac), when a time dependent
external magnetic field is applied parallel to the easy axis of the crystal.
We show that the back action of the cavity field on the sample significantly
increases the possibility of microwave emission. This radiation process can
be supperradiance or a maser-like effect, depending on the strength of the
dephasing. Our model provides further insight to the theoretical
understanding of the bursts of electromagnetic radiation observed in recent
experiments accompanying the resonant quantum tunneling of magnetization.
The experimental findings up to now can all be explained as being a maser
effect rather than superradiance. The results of our theory scale similarly
to the experimental findings, i.e., with increasing sweep rate of the
external magnetic field, the emission peaks are shifted towards higher field
values.
\end{abstract}

\maketitle

\section{Introduction}
Magnetic complex molecules have attracted a great deal of attention in recent
years, because they have remarkable properties, related to their high magnetic
anisotropy and large value of spin\cite{BTLCS99,WS99,V03}: for Mn$_{12}$Ac as
well as for Fe$_{8}$O, the quantum number of the total spin is $S=10.$
Accordingly, the eigenvalues of $S_{z}$, (the spin component in the direction
of the easy axis), can take $21$ different values: $m=-10,\ldots ,10.$ One of
the most interesting properties of these molecules is that in a slowly
changing external magnetic field the magnetization of the crystal consisting
of such molecules exhibit series of steps at sufficiently low
temperatures.\cite{FST96} The effect can be explained by assuming that the
energy levels of the molecules become doubly degenerate at the corresponding
values of the magnetic field and this resonance condition increases the
possibility of the
transition between the degenerate states with different values of $m$ and $%
m^{\prime }$. This kind of quantum tunneling between spin levels leads to a
sudden change in the magnetic moment of the crystal and is therefore of
fundamental importance as being a macroscopically observable quantum effect.

In an important theoretical work Chudnovsky and Garanin\cite{CG02} proposed
that resonant magnetic tunneling could be accompanied by the emission of
electromagnetic radiation, the possibility of superradiance from magnetic
molecules has been considered further in Ref.~[\onlinecite{HK03}] and bursts
of microwave pulses have actually been detected in recent
experiments.\cite{TCHA04,VSM04,HJAGHT05} According to
Refs.~[\onlinecite{CG02,HK03,JCC04}], the possible physical mechanism
responsible for this phenomenon can be superradiance (SR)\cite{D54,SRAD,GH82}
which is an interesting collective effect predicted first by Dicke in 1954,
and has been experimentally observed in several physical systems since
then.\cite{SRAD} However, when the radiation emitted by magnetic molecules was
detected, the sample was placed in a container, which acted as a waveguide.
This cavity changes the mode structure of the electromagnetic field
surrounding the sample, which is known to have crucial effects on the dynamics
of the emitted radiation. Studies in SR with other physical systems like the
ensemble of proton spins \cite{SRAD,KPSY88,BBZKMT90,YY04} in the MHz, and
with~Rydberg atoms in the GHz domain\cite{HR85} show that the presence of a
resonant cavity may enhance the collectivity of the radiating individual
dipoles, as first proposed by Bloembergen and Pound \cite{BP54}, and which
seems to be necessary to obtain radiation in the case of molecular magnets, as
well.\cite{YY05a,YY05b} Additionally, it has also been demonstrated that
external resonators such as Fabry-Perot mirrors can enhance the relaxation of
a crystal of molecular nanomagnets.\cite{TAH02} Inspired by these facts, we
investigate in this paper the interplay between the radiation and the changes
of the magnetization of a macroscopic sample of molecules Mn${}_{12}$-Ac
inside a nearly resonant cavity. We also note that the interesting proposal of
Ref.~[\onlinecite{LL01}] to use these molecules for implementing a quantum
algorithm gives another motivation to study their radiative properties.

The present paper is organized as follows: In Sec.~\ref{levelsection} we
investigate the relevant magnetic level structure and describe a method that
allows us to reduce the problem to a set of level pairs. The interaction of
the molecules with the cavity field is considered in
Sec.~\ref{interactionsection}. In Sec.~\ref{discussionsection} we discuss the
approximate analytical consequences of our model and present numerical results
as well. Finally we summarize and draw the conclusions
(Sec.~\ref{summarysection}).

\section{Magnetic level structure}

\label{levelsection} Experiments including magnetization measurements,
\cite{FST96,MSS01} neutron\cite{M99} and EPR\cite{BGS97,H98,HEJ03} studies on
crystals of Mn$_{12}$Ac and Fe$_{8}$O suggest that the Hamiltonian responsible
for the magnetic properties can be written as a sum of two terms:
\begin{equation}
H_{S}=H_{0}+H_{1}.  \label{H}
\end{equation}
Here $H_{0}$ is diagonal in the eigenbasis $\{|m\rangle \}$ of the
(dimensionless) $z$ component of the spin operator, $S_{z}$:
\begin{equation}
H_{0}=-DS_{z}^{2}-FS_{z}^{4}-\tilde{\mu}B_{0}S_{z},  \label{H0}
\end{equation}%
where the last term describes the coupling to an external magnetic field
applied in the $z$ direction (easy axis): $\mathbf{B}_{0}=(0,0,B_{0})$ with $%
\tilde{\mu}=g\mu _{B}$. On the other hand, $H_{1}$ consists of terms\cite%
{MSS01,BGS97} that do not commute with $S_{z}$:
\begin{equation}
H_{1}=C(S_{+}^{4}+S_{-}^{4})+E(S_{+}^{2}+S_{-}^{2})/2+K(S_{+}+S_{-})/2.
\label{H1}
\end{equation}%
As most of the experiments where microwave radiation emitted by magnetic
molecules was detected have been performed on Mn${}_{12}$-Ac, from now on,
we shall consider this molecule as the representative example. In this case
the values of the parameters in $H_{0}$ are $D/k_{B}=0.56K,$ and $%
F/k_{B}=1.1\cdot 10^{-3}K$. The coefficients in $H_{1}$ do not have
unanimously accepted values, but $H_{1}$ can be considered as a small
correction to $H_{0}$. However, as the transitions between levels with
different $m$ and $m^{\prime }$ are induced by terms that do not commute
with $S_{z}$, the importance of $H_{1}$ is fundamental from this point of
view. Tetragonal symmetry would allow only the quartic term, but there is
experimental evidence\cite{MSS01} showing the presence of weak quadratic and
linear terms in $H_{1}$. We shall return to the determination of the
coefficients in $H_{1}$ at the end of this section.

\begin{figure}[tbp]
\includegraphics[width=9 cm]{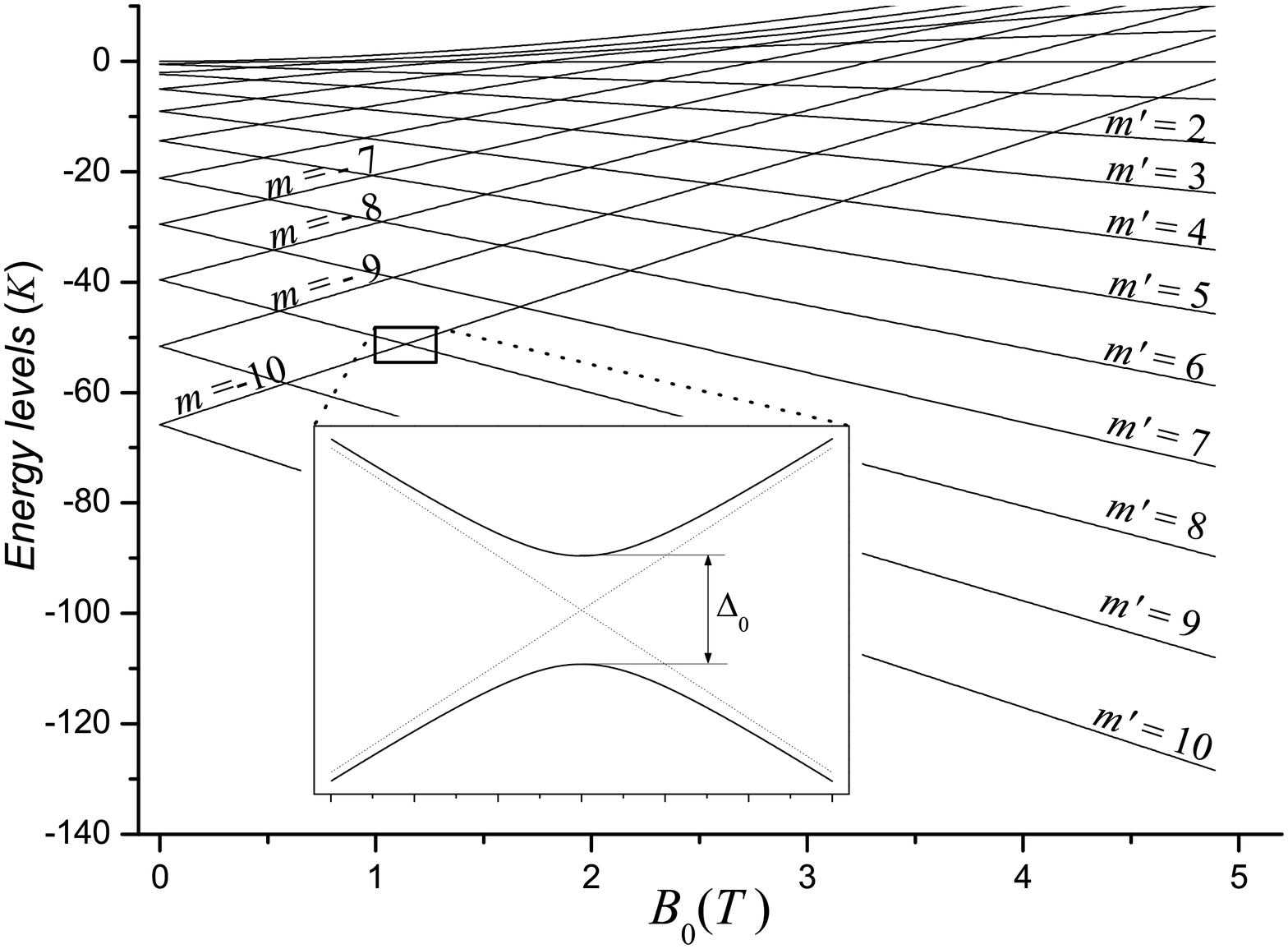}
\caption{The energy levels of Hamiltonian (\protect\ref{H}) as a function of
the external magnetic field $B_{0}$. The parameters are $D/k_{B}=0.56K,$ $%
F/k_{B}=1.1\cdot 10^{-3}K,$ $K=0.025\tilde{\protect\mu} B_{0}$, $%
E/k_{B}=-4.48$ $10^{-3}$ $K$ and $C/k_{B}=1.36$ $10^{-5}$ $K.$ The inset
zooms on the vicinity of $B_{0}=1.13$ T, where -- due to the presence of $%
H_{1}$ -- we can see a level splitting of $\Delta _{0}$ between the levels $%
m=-10$ and $m^{\prime }=8$.}
\label{levelscheme}
\end{figure}

In a typical experimental situation the external field $B_{0}$ slowly changes
in time, and consequently so does $H_{0}$. Considering the total Hamiltonian
(\ref{H}) as the generator of the time evolution (which means that relaxation
effects are not taken into account), the corresponding time dependent
Schr\"{o}dinger equation governs the dynamics. However, it turns out that its
solution is not feasible, because the time independent part of $H_{0}$ forces
a much faster evolution than the slow variation due to the change of the
magnetic field. The fundamental frequencies $\omega
_{mm^{\prime }}\approx (D/\hbar )(m^{\prime 2}-m^{2})$ are in the range $%
10^{10}-10^{11}s^{-1}$ being very fast compared with the time dependence of
the magnetic field $B_{0}$ that among ordinary circumstances\cite{TCHA04}
cause a change on the scale $10s^{-1}$. Therefore there is a significant
variation in the molecular state as a consequence of the first term in (\ref%
{H0}), while nothing happens due to $-\tilde{\mu}B_{0}S_{z}$, which appears
in the last term of $H_{0}$. On the other hand, we know that an appreciable
change in the state occurs only if two energy levels become close to each
other, as stipulated by a simple time dependent perturbation calculation,
where the energy difference between the levels appears in the denominator of
the transition probability. As the dominant term in the Hamiltonian (\ref{H}%
) is $H_{0}$, we can approximately find the points where two energy levels
are close to each other by calculating the eigenvalues of $H_{0}$, which are
obtained by the mere substitution $S_{z}\rightarrow m$. Simple algebra shows
that these eigenvalues become doubly degenerate with given $m$ and $%
m^{\prime }$ at the following values of $B_{0}$:
\begin{equation}
\tilde{\mu}B_{0}=-D(m+m^{\prime })\left( 1+\frac{F}{D}\left[ m^{2}+m^{\prime
2}\right] \right) .  \label{cross}
\end{equation}%
A part of the level scheme of the total Hamiltonian $H_{S}$ is shown in Fig.~%
\ref{levelscheme} as function of $B_{0}$. The special values of $B_{0}$
given by Eq.~(\ref{cross}) -- where the levels of $H_{0}$ cross -- can be
clearly identified in this figure. However, as it is known, the presence of $%
H_{1}$ perturbs the eigenvalues leading to a splitting of the levels instead
of crossing as shown by the inset. The resonance condition implies that
appreciable changes in the population of the levels is expected around these
avoided crossings (sometimes also called anticrossings). Thus the system can
be efficiently approximated by a set of level pairs, each of which is to be
considered as an effective two-level system, similarly to the figure shown
in the inset.
%%%%%%%%%%%%%%%%%%%%%%%%%%%%%%%%%%%%%%%%%%%%%%%%%%%%%%%%%%%%%%%%%%%%%%%%%%%
%%%%%%%%%%%%%%%%%%%%%%%%%%%%%%%%%%%%%%%%%%%%%%%%%%%%%%%%%%%%%%%%%%%%%%%%%%%

Reducing the problem to a set of level pairs means technically that one
applies degenerate perturbation theory around each avoided level crossing
(determined by the values of $B_{0}$ in Eq.~(\ref{cross})) following a
technique proposed independently by van Vleck\cite{vV29}, Des Cloiseaux\cite%
{dC60} and other authors as summarized in Ref.~[\onlinecite{K74}]. In the
vicinity of a given avoided level crossing we find a unitary transformation
that block diagonalizes $H_{S}$ on the zero order two-dimensional
eigensubspaces and generate an effective Hamiltonian $H_{e}$ that has the
same eigenvalues as $H_{S}$:
\begin{equation}
H_{e}=UH_{S}U^{\dagger }  \label{effham}
\end{equation}%
with
\begin{equation}
U=\sum_{a}(P_{a}^{0}P_{a}P_{a}^{0})^{-1/2}P_{a}^{0}P_{a},  \label{unitary}
\end{equation}%
where $a$ labels the various eigensubspaces with quasi-degenerate
eigenvalues, and $P_{a}^{0}$ are orthogonal projections on the degenerate
eigensubspaces of $H_{0}$. For a given value of $a,$ corresponding to the
avoided crossing of levels $m$ and $m^{\prime }$, we have $P^{0}=|m\rangle
\langle m|+|m^{\prime }\rangle \langle m^{\prime }|.$ As the perturbation is
turned on, the $\{\left\vert m\right\rangle ,\left\vert m^{\prime
}\right\rangle \}$ eigenstates of $H_{0}$ evolve into $\{\left\vert \phi
_{m}\right\rangle ,\left\vert \phi _{m^{\prime }}\right\rangle \}$ and we
consider $P=|\phi _{m}\rangle \langle \phi _{m}|+|\phi _{m^{\prime }}\rangle
\langle \phi _{m^{\prime }}|$ which projects onto the space arising from the
zero order subspace of interest. By appropriate expansions\cite%
{G91,LL00,YP05} of this unitary operator $U$, one can obtain a perturbation
series, leading to useful analytical approximations of $H_{e}.$
Alternatively, the numerically exact projections $P_{a}$ can be found by the
diagonalization of $H_{S}$, and then Eqs.~(\ref{effham}) and (\ref{unitary})
provide the operator $H_{e}$. The latter method is followed in this paper
and we obtain in each two-dimensional subspace spanned by $\{\left\vert
m\right\rangle ,\left\vert m^{\prime }\right\rangle \}$ the following matrix
for the effective Hamiltonian:
\begin{equation}
H_{e}(t)=\left(
\begin{array}{cc}
\varepsilon _{0}+w/2 & \Delta _{0}/2 \\
\Delta _{0}/2 & \varepsilon _{0}-w/2%
\end{array}%
\right) _{m,m^{\prime }}  \label{He}
\end{equation}%
with time dependent elements, and of course the values depend also on the
pair $\{\left\vert m\right\rangle ,\left\vert m^{\prime }\right\rangle \}$.
Here $\varepsilon _{0}$ is the energy where the given crossing would occur, $%
w$ is proportional to the time dependent external field in the $z$
direction, while the offdiagonal element $\Delta _{0}$ is the level
splitting responsible for the effective coupling between the levels. $w$
will be assumed to be linear in time with constant $\dot{B}_{0}$, yielding $%
w(t)=-\tilde{\mu}\dot{B}_{0}(t-t_{0})(m-m^{\prime })$ with $t_{0}$ being the
time instant when the crossing would occur. Note that this is a reasonable
approximation even for a time scale much longer than the expected duration of
the transition to be described (see e.g.~Fig.~1.~in
Ref.~[\onlinecite{VSM04}]). This linear approximation corresponds to the usual
Landau-Zener-St\"{u}ckelberg (LZS) model \cite{L32,Z32,S32}, by the aid of
which we can calculate the probability of a given $m\rightarrow m^{\prime }$
transition: $P_{mm^{\prime }}=1-\exp (-\pi \Delta _{0}^{2}/2\hbar w)$. Note
that in this expression both $w$ and $\Delta _{0}$ depend on the labels $%
m,m^{\prime }$. For a given pair of levels and sweep rate $\dot{B}_{0}$, $%
P_{m,m^{\prime }}$ is determined by the magnitude of the level splitting $%
\Delta _{0}$, i.e., essentially by the parameters in $H_{1}$. If we assume
that initially the system is in thermal equilibrium, we can consider a
series of transitions at the values of $B_{0}$ given by Eq.~(\ref{cross}).
Calculating the expectation value of the operator $S_{z}$ (which is
proportional to the magnetization) we obtain a staircase-like hysteresis
loop that can be compared with the experimental curves\cite{MSS01} at a
given temperature and sweep rate. As our results depend on the coefficients
in $H_{1}$, the minimization of the difference between the steps in the
calculated hysteresis curve and the experimental plots gives the desired
parameter values. We have found the best agreement for $K=0.025\tilde{\mu}%
B_{0}$, $E/k_{B}=-4.48$ $10^{-3}$ $K$, $C/k_{B}=1.36$ $10^{-5}$ $K$,
therefore these parameter values will be used in the following.

The method summarized in this section, first of all, gives us information
about the magnitude of the terms in the spin Hamiltonian, and describes how
to obtain the level splittings $\Delta _{0}$. As we shall see in the next
section, the coupling of the molecular system to the cavity field at a given
$m\rightarrow m^{\prime }$ transition can also be determined in this way.

\section{Interaction with the resonant cavity field}

\label{interactionsection} In this section we describe the interaction of an
ensemble of magnetic molecules with a quasi-resonant cavity field. The time
dependence of the level structure, considered in the previous section, will
bring a certain level pair into resonance with the cavity field at a given
value of the external magnetic field $\mathbf{B}_{0}.$ Usually there are
avoided level crossings before the resonance, where for low lying $m$ and
$m^{\prime}$ the LZS transition probabilities $P_{m,m^{\prime }}$ are small,
leading to an almost complete inversion. We shall consider a dynamical
equation for the density operator of the molecules and assume that a dipole
moment is generated during a certain transition between the split levels $m$
and $m^{\prime }$ of a given molecule which in turn serves as a source of
microwave radiation influencing the transitions in all other molecules. To
take into account this radiation mediated interaction, we have to add a new
term to the effective two-level Hamiltonian describing
the interaction with the magnetic dipole field of the cavity, $\mathcal{\vec{%
B}}$, which is an additional field beyond the stronger but almost static
magnetic field $\mathbf{B}_{0}$ creating the inversion between the levels.
The fast dynamics of the magnetic dipoles will be a forced oscillation
generated by the interaction with the external field that is characterized
by the interaction Hamiltonian $H_{I}=\tilde{\mu}\mathcal{\vec{B}}\mathbf{%
\tilde{S}}_{e}.$ Here $\mathbf{\tilde{S}}_{e}=USU^{\dagger }=(\tilde{S}_{x},%
\tilde{S}_{y},\tilde{S}_{z})$ denotes the spin operator one obtains after
the unitary transformation (\ref{effham}) described in the previous section,
by restricting it to the actual two-dimensional subspace we consider. The
matrix elements of $\mathbf{\tilde{S}}_{e}$ usually differ from those of $%
\mathbf{S.}$ As for the field strength $\mathcal{\vec{B}},$ a detailed model
should take into account the mode structure of the cavity. However, the
characteristic features of the dynamics due to the cavity field can be
captured by a simpler model to be discussed here. We consider the microwave
field as a single transverse (TM) mode being perpendicular to the $z$ (easy)
axis and having a frequency $\Omega $ and equal amplitudes in the $x$ and $y$
directions:
$H_{I}=-\tilde{\mu}\mathcal{B}(\tilde{S}_{x}+\tilde{S}_{y})/\sqrt{2}$. We note
that the choice of the polarization does not have essential influence on the
results presented here. Now the equations describing the dynamics of the
two-level system (without relaxation) can be written as
\begin{equation}
\frac{\partial \varrho }{\partial t}=-\frac{i}{\hbar }\left[ H^{\prime
},\varrho \right] ,  \label{master}
\end{equation}%
where $\varrho $ is the density operator of the effective two-level system
and
\begin{equation}
H^{\prime }=H_{e}+H_{I}=\left(
\begin{array}{cc}
\varepsilon _{0}+\hbar \omega /2 & \Delta /2 \\
\Delta ^{\ast }/2 & \varepsilon _{0}-\hbar \omega /2%
\end{array}%
\right) _{mm^{\prime }}.
\end{equation}%
Here $\hbar \omega (t)=w(t)-2\tilde{\mu}\mathcal{B}s^{\prime }$ and $\Delta
=\Delta _{0}-2\tilde{\mu}\mathcal{B}s$ with $s^{\prime }$ and $s$ being the
diagonal and offdiagonal elements resulting from the coupling operator
$(\tilde{S}_{x}+\tilde{S}_{y})/\sqrt{2}$. This leads to the following
equations for the population differences and the coherences between the
states:
\begin{eqnarray}
\frac{d}{dt}(\varrho _{mm}-\varrho _{m^{\prime }m^{\prime }}) &=&\frac{i}{%
\hbar }(\Delta ^{\ast }\varrho _{mm^{\prime }}-\Delta \varrho _{m^{\prime
}m})  \label{denm1} \\
\frac{d}{dt}\varrho _{mm^{\prime }} &=&-i\omega (t)\varrho _{mm^{\prime }}+%
\frac{i}{2\hbar }\Delta (\varrho _{mm}-\varrho _{m^{\prime }m^{\prime }}).
\end{eqnarray}%
As $\Omega $ is in the terahertz domain, we can separate a slowly varying
amplitude of the time varying field and write
\begin{equation}
\mathcal{B}=\left( \frac{1}{2}B(t)e^{-i\Omega t}+c.c.\right) u_{k}(z),
\label{Bslow}
\end{equation}%
where $u_{k}(z)$ is the corresponding mode function of the cavity, and $%
\left\vert \dot{B}(t)\right\vert \ll \Omega \left\vert B\right\vert $. This
makes straightforward a similar separation for the offdiagonal elements of
the density matrix: $\varrho _{mm^{\prime }}=R_{mm^{\prime }}e^{-i\Omega t}$
where $R_{mm^{\prime }}$ is again assumed to vary slowly compared with $%
e^{-i\Omega t}.$ Substituting into Eq.~(\ref{denm1}) we can neglect terms
oscillating with frequency $2\Omega $ as they do not contribute essentially
to the evolution of the state, i.e, this is the standard rotating wave
approximation (RWA)\cite{SSL74}. Introducing the notation $Z_{mm^{\prime
}}=\varrho _{mm}-\varrho _{m^{\prime }m^{\prime }}$ for the inversion
between levels $m$ and $m^{\prime }$, we obtain:
\begin{eqnarray}
\frac{d}{dt}Z_{mm^{\prime }} &=&\frac{i}{2\hbar }\left\{ \left(
R_{mm^{\prime }}\Delta _{0}^{\ast }e^{-i\Omega t}-R_{mm^{\prime }}^{\ast
}\Delta _{0}e^{i\Omega t}\right) -\tilde{\mu}(B^{\ast }s_{mm^{\prime
}}^{\ast }R_{mm^{\prime }}-Bs_{mm^{\prime }}R_{mm^{\prime }}^{\ast
})u_{k}^{2}(z)\right\}  \label{Zd} \\
\frac{d}{dt}R_{mm^{\prime }} &=&-i(\omega (t)-\Omega )R_{mm^{\prime }}+\frac{%
i}{\hbar }(\Delta _{0}e^{i\Omega t}-\tilde{\mu}Bs_{mm^{\prime
}})Z_{mm^{\prime }}\text{.}  \label{Rd}
\end{eqnarray}

At a given time instant only one of the level pairs get into resonance with
the cavity, therefore from now on we shall omit the indices $m,m^{\prime }$.
We shall come back to this point, and discuss the mechanism which selects
the actual level pair. We can also average out the equations over a time
period of a cycle of the oscillation that eliminates the terms varying with
frequency $\Omega $. A similar procedure can be performed in space over the
wavelength $\lambda =2\pi /k$, and we shall make use of $\frac{1}{\lambda }%
\int_{0}^{\lambda }u_{k}^{2}(z)=1/2$.

We also have to describe the effects caused by other degrees of freedom.
These additional interactions -- among which the strongest one is the spin
phonon coupling, i.e, the oscillation of the atoms in the lattice -- are not
taken into account by the Hamiltonian (\ref{H}), but can significantly
influence the dynamics. The effects due to the reservoir of phonons (i) can
be dissipative, by simply taking up the energy from the spin system and (ii)
dephasing, by randomly disturbing the relative phases of the magnetic
states. Dissipative terms lead to a decay of the diagonal elements while
dephasing reduce the off-diagonal elements of $\mathcal{\varrho }$. The
diagonal terms relax generally much slower than the offdiagonal ones,
therefore we consider only this (so called transversal) relaxation. As
usual, it will be taken into account by assuming a simple exponential decay
with a time constant $T_{2}$, the order of magnitude of which can be
estimated between $10^{-5}$ and $10^{-7}$ \emph{s} in the temperature range
we are interested in\cite{LL00}. With this term we have:
\begin{eqnarray}
\frac{d}{dt}Z &=&-\frac{i\tilde{\mu}}{2\hbar }(B^{\ast }s^{\ast }R-BsR^{\ast
}),  \label{Bl1} \\
\frac{d}{dt}R &=&-i(\omega (t)-\Omega )R-\frac{i}{2\hbar }\tilde{\mu}%
BsZ-R/T_{2}.  \label{Bl2}
\end{eqnarray}%
These equations are familiar from the theory of magnetic resonance, and as
it is known, the effect of the field $B$ is essential when the frequency
originating from the slowly varying longitudinal field gets close to the
cavity frequency: $\omega (t)\approx \Omega $.

\bigskip We also treat the mode amplitude of the cavity as a dynamical
quantity. Following the usual semiclassical approach of radiation-matter
interaction theory\cite{SSL74}, the time varying field resulting from the
magnetic molecules of the crystal will be described here as the field of a
sample with time dependent magnetic dipole moment density $\mathcal{M}$ in
the $x-y$ plane. The appropriate component of the transverse $\mathcal{H}$
field originating from $\mathcal{M}$ as a source, obeys the damped
inhomogeneous wave equation
\begin{equation}
\Delta \mathcal{H}-\mathcal{\dot{H}}/(c^{2}T_{c})\mathcal{-}\ddot{\mathcal{H}%
}/c^{2}=\ddot{\mathcal{M}}/c^{2},  \label{wave}
\end{equation}%
where $T_{c}$ is the cavity lifetime. Within the cavity we expand the field
into modes, and in accordance with Eq.~(\ref{Bslow}), we also write:%
\begin{equation}
\mathcal{H=}\left( \frac{1}{2}He^{-i\Omega t}+c.c.\right) u_{k}(z),\quad
\mathcal{M=}\left( \frac{1}{2}Me^{-i\Omega t}+c.c.\right) f(z)  \label{mfast}
\end{equation}%
where $f(z)$ is nonzero only within the sample, where it can be taken equal to
$u_{k}(z)$. If one substitutes into Eq.~(\ref{wave}), and makes an
approximation exploiting that the amplitudes, $H$ and $M$ are slowly varying%
\cite{SSL74} with respect to $e^{-i\Omega t}$, one obtains the equation
\begin{equation}
\frac{dH}{dt}=i\Omega \eta \frac{M}{2}-\frac{1}{2T_{c}}H.
\end{equation}%
The filling factor
\begin{equation}
\eta =\int_{\mathcal{C}}u_{k}(z)f(z)dz/\int_{\mathcal{C}}u_{k}^{2}(z)dz%
\approx l/L
\end{equation}%
arises when we project the resulting equation on the mode in question, by
integrating over the volume of the cavity. Here $l/L$ is the ratio of the
lengths of the sample and the cavity, corresponding to the geometries
reported in the experiments\cite{TCHA04,VSM04}.

The corresponding component of the transverse magnetization of the sample is
given by
\begin{equation}
\mathcal{M}=N_{0}\tilde{\mu}Tr\left[\varrho
(\tilde{S}_{x}+\tilde{S}_{y})/\sqrt{2}\right]=N_{0}\tilde{\mu} (s^{\ast
}\varrho _{mm^{\prime }}+s\varrho _{m^{\prime }m}),  \label{Mcal}
\end{equation}%
where $N_{0}$ is the number density of the molecules participating in the
transition $m\rightarrow m^{\prime }$. We note that the static part of $%
\mathcal{M}$ containing $s^{\prime },$ does not give rise to radiation. Eq.~(%
\ref{Mcal}) connects the microscopic dynamical variables with the
macroscopic ones. Recalling that $\varrho _{mm^{\prime }}=R_{mm^{\prime
}}e^{-i\Omega t}$, we see that $M=2N_{0}\tilde{\mu}s^{\ast }R_{mm^{\prime
}}. $ The slowly varying magnetic induction field acting on the molecules is
given by $B=\mu _{0}(H+\beta M)$, where $\mu _{0}$ is the vacuum
permeability and $\beta $ may differ from unity giving account of a local
field correction resulting from the near field of the dipoles\cite{J98}. It
is natural to measure the time variable in units of the characteristic time
\begin{equation}
T_{0}=\left( \frac{2\hbar }{\eta N_{0}\Omega \mu _{0}\tilde{\mu}%
^{2}\left\vert s\right\vert ^{2}}\right) ^{1/2}.  \label{T0def}
\end{equation}%
As we shall see, the relation of $T_{0}$ and the rate of relaxation
characterizes the dynamics: in case of $T_{0}/T_{2}\ll 1$ the phase
correlation of the individual emitters is conserved during the process and a
superradiant pulse (or a sequence of pulses) can be emitted. On the other
hand, $T_{0}/T_{2}>1$ indicates that relaxation effects are too strong to
allow SR to occur.

To perform the calculations it is straightforward to introduce the
dimensionless magnetic field strength and induction amplitudes:
\begin{equation}
h=\left( \frac{\mu _{0}}{N_{0}\hbar \Omega }\right) ^{1/2}H,\quad \text{ \ }%
b=\left( \frac{1}{\mu _{0}N_{0}\hbar \Omega }\right) ^{1/2}B.
\end{equation}%
The field intensity can be measured as the energy density averaged out over
the time and space period. The dimensionless field intensity $I=\left\vert
h\right\vert ^{2}/2$ gives the number of emitted photons of energy $\hbar
\Omega $ per number of molecules participating in the given transition.
Outside the sample $h=b,$ while within the sample one has
\begin{equation}
b=h+2\beta \frac{1}{\eta T_{0}\Omega }Re^{-i\psi },  \label{nodifference}
\end{equation}%
where $\psi $ is the phase of the offdiagonal coupling constant: $%
s=|s|e^{i\psi }$. The dynamical equation for the magnetic field in
dimensionless form reads:
\begin{equation}
\frac{dh}{d\tau }=-\frac{\kappa }{2}h+iRe^{-i\psi },  \label{hp}
\end{equation}%
where $\tau =t/T_{0}$ and $\kappa =(T_{0}/T_{c}),$ is the damping
coefficient of the cavity. These equations are to be solved together with
Eqs.~(\ref{Bl1},\ref{Bl2}), which take the dimensionless form:
\begin{eqnarray}
\frac{d}{d\tau }Z &=&-i(b^{\ast }Re^{-i\psi }-bR^{\ast }e^{i\psi }),
\label{Zp} \\
\frac{d}{d\tau }R &=&-iT_{0}(\omega (\tau )-\Omega )R-ibe^{i\psi }Z-\gamma R.
\label{Rp}
\end{eqnarray}%
with $\gamma =T_{0}/T_{2}$. Note that numerically (using SI units)
$T_{0}\approx |s|^{-1}\times 10^{-8}s$, where we substituted
$N_{0}\eta=10^{23}m^{-3}$, corresponding to the values reported in the
experiment\cite{TCHA04}. Depending on the transition $m\rightarrow m^{\prime
}$, usually the magnitude of the dimensionless matrix element $|s|$ is much
less than unity, thus $10ns$ (obtained with $|s|=1$) is basically a lower
bound for $T_{0}$. This value -- at least at low temperatures -- is less than
the time scale of the relaxation, $T_{2}$, but clearly by orders of magnitude
larger than the period of the microwave
radiation $1/\Omega $, thus our rotating wave approximation leading to Eqs.~(%
\ref{Bl1},\ref{Bl2}) is valid. Additionally, as $\Omega $ is around $10^{11}$
$s^{-1}$, Eq.~(\ref{nodifference}) shows that it is a very good assumption
to take $h=b$ within the sample as well.

%%%%%%%%%%%%%%%%%%%%%%%%%%%%%%%%%%%%%%%%%%%%%%%%%%%%%%%%%%%%%%%%%%%%%%

\section{Results and Discussion}

\label{discussionsection} The generic scheme for the emission from the
ensemble of magnetic molecules considered in this paper starts with inverted
two-level systems that come into resonance with the cavity field at a certain
value of the external magnetic field $B_{0}$. Besides resonance, an additional
requirement for the transverse radiation to begin is that the wavelength
corresponding to the transition frequency should be comparable or smaller than
the size of the sample, otherwise the non-transverse near-field of the sources
would dominate the emitted field at the location of the other molecules. This
explains why in the experiments reported in Ref.~[\onlinecite{TCHA04}] the
emission is in the $mm$ range wavelength.

Now we shall analyze if the observed radiation can be considered as
superradiance demanding $\gamma \ll 1$, or is it rather a maser effect, where
the absence of phase relaxation is not crucial. Therefore we first assume that
$\gamma = 0,$ and \ see that equations (\ref{Zp}) and (\ref{Rp}) admit a
simple constant of motion:
\begin{equation}
Z^{2}+2\left\vert R\right\vert ^{2}=Z_{0}^{2}.
\end{equation}%
If $\omega (\tau )$ is changing sufficiently slowly, the condition of
resonance $\omega (\tau )-\Omega _{c}=0$ is sustained during the dynamics of
the emission. Then writing $Z=Z_{0}\cos \theta (\tau )$, and $\left\vert
R\right\vert =(Z_{0}/\sqrt{2})\sin \theta (\tau )$, a simple equation yielding
essential physical insight into the nature of the problem can be obtained.
With the assumptions that $R$ is real, $\psi =\pi /2$ and using
Eqs.~(\ref{Zp},\ref{Rp}), one has $b=\frac{d}{d\tau }\theta (\tau
)/\sqrt{2}\equiv \dot{\theta}/\sqrt{2}$, and from Eq.~(\ref{hp}) we obtain
\begin{equation}
\ddot{\theta}(\tau )+\kappa \dot{\theta}(\tau )/2-Z_{0}\sin \theta (\tau )=0,
\label{pendulum}
\end{equation}%
which is the equation of a damped pendulum ($\theta $ measured from the
inverted position), being often discussed in coherent atom-field
interactions. The physically realistic initial condition for $\theta _{0}$
is a small ($\theta _{0}\approx 0$) value, as initially we expect the
offdiagonal element of the density matrix to be small. This comes from the
small initial polarization as a remnant of the coherence between the levels
during the magnetic tunneling transition. Putting in such an initial
condition takes into account the rapidly varying terms omitted in Eqs.~(\ref%
{Rd}) and (\ref{Zd}) which would also lead to a small but nonzero $R$
initially. Experimental results reported in Ref.~[\onlinecite{TCHA04}]
suggest that the radiation appears after a crossing with small tunneling
probability $P_{mm^{\prime }}$, thus there is a significant inversion
present in the system, meaning $Z_{0}\approx 1$ in the beginning of the
radiation process.

If the cavity lifetime is short, $\kappa $ is large, the second term dominates
over the first in Eq.~(\ref{pendulum}). Neglecting $\ddot{\theta},$ the
equation of the overdamped pendulum admits an analytical solution. For the
emitted intensity one obtains
\begin{equation}
\left\vert b\right\vert ^{2}=\dot{\theta}^{2}(\tau )/2=\frac{1}{2\tau
_{R}^{2}}\mathrm{sech}^{2}[(\tau -\tau _{d})/\tau _{R}],  \label{intens}
\end{equation}%
where $\tau _{R}=\frac{\kappa }{2Z_{0}}$, and $\tau _{d}=\tau _{R}(\ln
\theta _{0}/2)$. In this bad cavity limit $T_{c}$ can be estimated as $L/c,\
$the time needed for a photon to leave the cavity of length $L$. Assuming $%
Z_{0}=1$, the characteristic time of the emission in usual units is given by
$T_{R}=T_{0}\tau _{R}=T_{0}^{2}\frac{c}{2L}=\frac{\hbar c}{\eta LN_{0}\Omega
_{c}\mu _{0}\tilde{\mu}^{2}\left\vert s\right\vert ^{2}}$. We see that this
time constant is inversely proportional to the number density of the
molecules, $N_{0}$, while according to Eq.~(\ref{intens}), the intensity of
the radiation is proportional to $N_{0}^{2}$: these are the characteristic
features of superradiance \cite{SRAD}. In addition, there is a delay time $%
T_{d}=\tau _{d}T_{0}$ necessary for the appearance of the pulse described in
Eq.~(\ref{intens}).

In the case the cavity losses do not dominate the process, the energy of the
field is fed back into the crystal and according to Eq.~(\ref{pendulum})
this leads to a damped periodic process. Assuming a perfect cavity one has a
kind of Rabi oscillations with a time dependent field: energy is exchanged
periodically between the crystal and the field.

These considerations based on the analytic solutions, however, become only
valid approximately, as they do not take into account the factor $\omega
(\tau )-\Omega _{c}$ in Eq.~(\ref{Rp}). As we assume a constant $\dot{B}_{0}
$, we can introduce a constant dimensionless external field sweep rate $v$
via
\begin{equation}
T_{0}(\omega (\tau )-\Omega _{c})=v\tau ,
\end{equation}%
where the origin of the time axis is chosen so that $\tau =0$ corresponds to
exact resonance: $\omega (0 )=\Omega _{c}$. As an example, at the
$m=-10\rightarrow m^{\prime }=8$ transition with $\dot{B}_{0}=30$ mT/s and
$T_{0}=10^{-6} s$ we have $v=T_{0}^{2}\tilde{\mu}\dot{B}_{0}(m-m^{\prime
})/\hbar \approx 0.1$, leading to a dynamics significantly different from the
analytical solution.
%%%%%%%%%%%%%%%%%%%%%%%%%%%%%%%%%%%%%%%%%%%%%%%%%%%%%%%%%%%%%%%%%%%%%%%%%%%
\begin{figure}[tbp]
\includegraphics[width=9 cm]{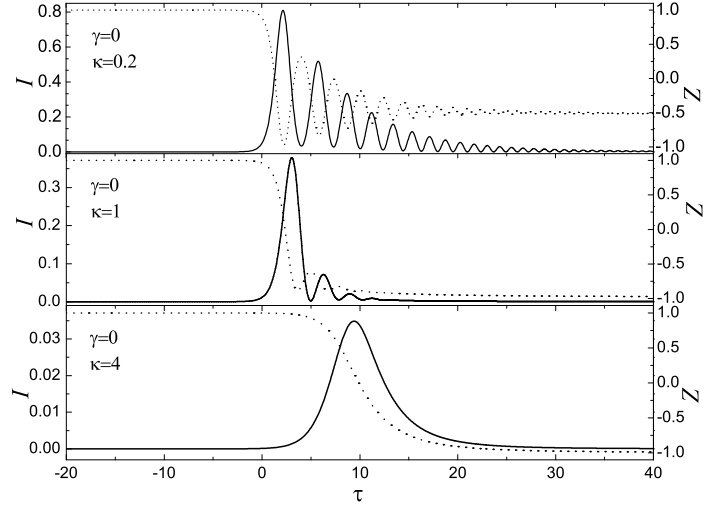}
\caption{Population difference $Z$ (dotted line) and the dimensionless
intensity (solid line) of the radiation emitted by the molecular system as a
function of $\protect\tau $
for different cavity decay rates. The transversal relaxation rate $\protect%
\gamma =0$ and the dimensionless sweep rate corresponding to these plots is $%
v=0.2$. Note that the energy per unit volume that corresponds to the field $%
I=1$ means that each active molecule in the sample emits a photon of energy $%
\hbar \Omega.$} \label{SRfig}
\end{figure}
%%%%%%%%%%%%%%%%%%%%%%%%%%%%%%%%%%%%%%%%%%%%%%%%%%%%%%%%%%%%%%%%%%%%%%%%%%%
Qualitatively, we expect that around $\tau =0$ (crossing point) the coupling
begins to act, and creates a superposition of the levels. The coherence of
the levels begin to increase accompanied by a finite transition probability:
$Z$ will differ substantially from $Z(-\infty ).$ Then the oscillation of
the pendulum, i.e., the radiation starts, but as the levels separate, their
energy difference and therefore the oscillation frequency becomes larger.
Taking relaxation into account, the amplitude of these oscillations
diminishes, the molecules do not emit radiation any longer and
simultaneously $Z$ will reach a stationary value $Z(\infty ),$ analogously
to what is usually called quantum tunneling of magnetization, because a
different $Z$ means different value of the expectation value of $S_{z}.$ In
this sense the inclusion of this time dependent detuning leads to a similar
effect as discussed in the problem of tunneling, but the resonance condition
is ensured by the inclusion of the time dependent cavity field, therefore
the dynamics is more complicated than in LZS theory.

Quantitatively, Fig.~\ref{SRfig} shows the inversion and the intensity of the
emitted radiation as a function of time for different strengths of the cavity
decay. The effect of the resonance is clear, appreciable radiation and change
in $Z$ is seen after $\tau =0$. As the cavity damping becomes stronger we have
less oscillations in the emitted intensity and the process starts later. For
small values of $\kappa $ the final inversion $Z(\infty )$ is determined by
the sweep rate, but when cavity losses become significant, the energy of the
molecular system is lost via the cavity field during the process leading to
$Z(\infty )\approx -1$. A bad cavity ($\kappa \gg 1$) overdamps the pendulum
and the system leaves the vicinity of the resonance before observable emission
occurs. We note that the inversion $Z(\infty )$ is not necessarily closely
related to the final magnetization of the sample, because following the photon
emission, cascade (not purely two-level) transitions related to a given side
of the two-well potential can also have a considerable probability.
%%%%%%%%%%%%%%%%%%%%%%%%%%%%%%%%%%%%%%%%%%%%%%%%%%%%%%%%%%%%%%%%%%%%%%%%%%%
\begin{figure}[tbp]
\includegraphics[width=9 cm]{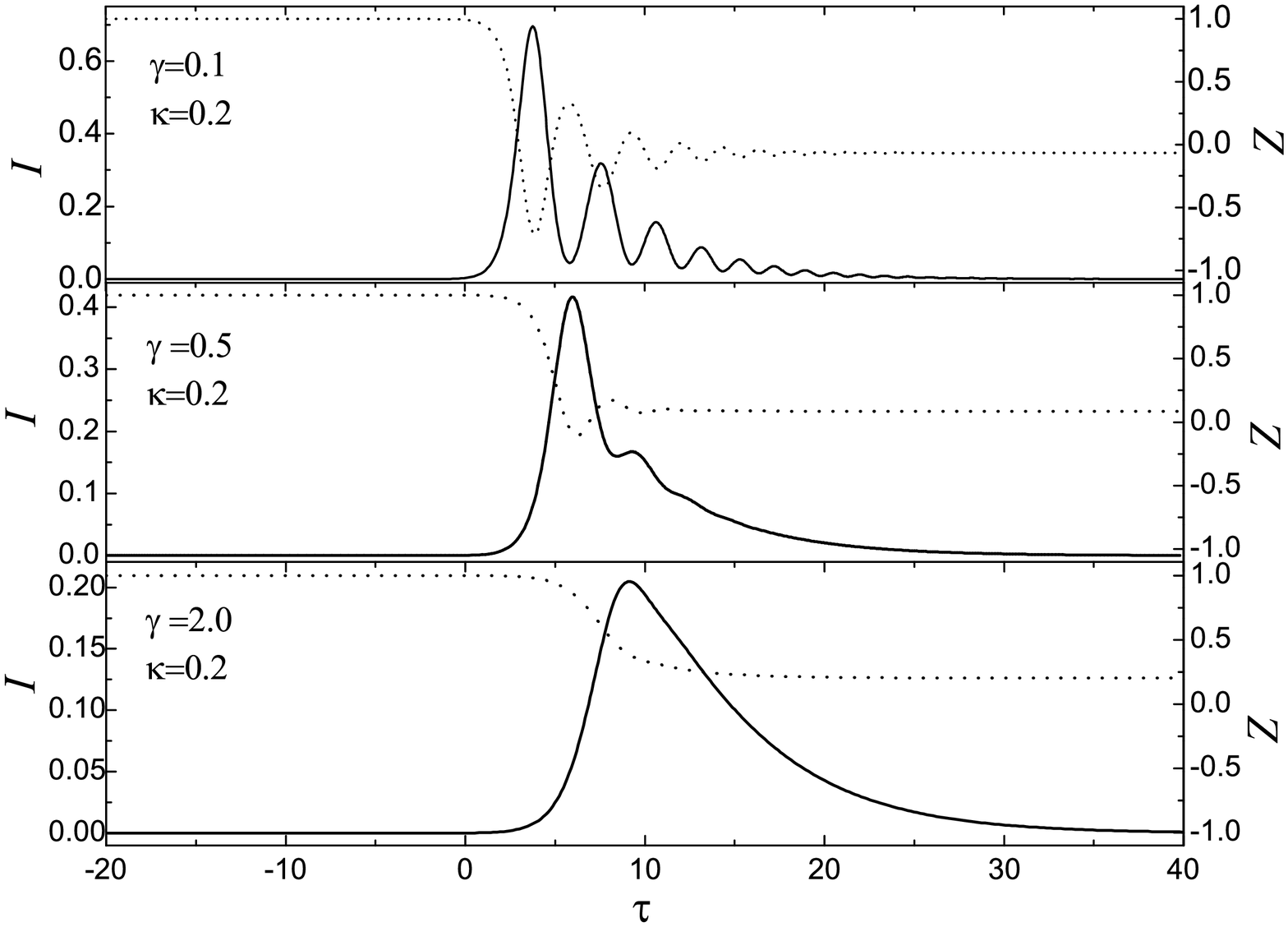}
\caption{The effects of the transversal relaxation on the dynamics of the
population difference $Z$ (dotted line) and the intensity (solid line) of the
radiation emitted by the molecular system. In comparison to
Fig.~\protect\ref{SRfig}, note the saturation effect in $Z$: for a given sweep
rate (in this case $v=0.2$) increasing values of $\protect\gamma $ push the
final population difference towards positive values.} \label{gammafig}
\end{figure}
%%%%%%%%%%%%%%%%%%%%%%%%%%%%%%%%%%%%%%%%%%%%%%%%%%%%%%%%%%%%%%%%%%%%%%%%%%%

\bigskip
So far it has been assumed that the phase memory of the system is conserved,
$\gamma =T_{0}/T_{2}$ is small and accordingly the damping term $-\gamma R$
was neglected in Eq.~(\ref{Rp}). This was the assumption that led to the
coherent behavior of the molecules resulting in superradiant emission.
However, in reality there are at least two reasons to consider nonzero
$\gamma$. One of them is the spin-phonon coupling which is temperature
dependent, thus can be reduced by cooling the sample. Additionally, if the
size of the system is smaller than the wavelength the near field dipole-dipole
coupling between the molecules becomes important and can be shown to lead to
an effective  phase relaxation.\cite{GH82,FHM73} Microscopic studies of the
latter effect can be found in Refs.~[\onlinecite{SRAD,ZMT83}], for a detailed
recent work see Ref.~[\onlinecite{DHTK05}]. At very low temperatures this
effect can be even stronger than the homogeneous broadening mechanism caused
by elastic collisions with the phonons. At temperatures around $2$K, however,
where the experiments observing the radiation have been performed, the
dephasing is predominantly due to spin-phonon interactions instead of
dipole-dipole coupling.\cite{Se95,LL99} In the present work all these
relaxation mechanisms are incorporated effectively by an appropriately chosen
damping coefficient $\gamma $.

The consequences of phase relaxation is shown in Fig.~\ref{gammafig}, where a
moderate constant cavity decay ($\kappa =0.2$) is also taken into account. For
weak dephasing, we have similar oscillations in $Z$ and pulse structure as
shown in Fig.~\ref{SRfig}. Increasing the value of $\gamma$ the
coherent Rabi oscillations disappear. Additionally, the final inversion $%
Z(\infty )$ is a monotonically increasing function of $\gamma $, and this can
be considered as a remarkable difference between the two decay mechanisms.
Note that this saturation effect can be responsible for the additional steps
in the hysteresis curve published in Ref.~[\onlinecite{TCHA04}] following the
most pronounced one which is accompanied by microwave radiation: The nonzero
population that remains on the upper level after the transition can lead to an
observable change of the magnetization of the sample at a next avoided level
crossing.

In the case of strong dephasing, the time derivative of $R$ can be neglected
with respect of $\gamma R,$ and from Eq.~(\ref{Rp}) we obtain that $R$
follows adiabatically the time dependence of $b$. Substituting back into
Eq.~(\ref{Zp}) we obtain the following rate equation description of the
process:
\begin{eqnarray}
\frac{dZ}{d\tau } &=&-Z\left\vert b\right\vert ^{2}\frac{2\gamma }{\gamma
^{2}+v^{2}\tau ^{2}},  \notag \\
\frac{db}{d\tau } &=&-\frac{\kappa }{2}b+\frac{bZ}{\gamma +iv\tau }.
\label{rate}
\end{eqnarray}%
Here atomic coherence does not play any role, thus the process cannot be
termed as superradiance, it is rather a maser, operating on the inverted
magnetic levels.

\bigskip
%%%%%%%%%%%%%%%%%%%%%%%%%%%%%%%%%%%%%%%%%%%%%%%%%%%%%%%%%%%%%%%%%%%%%%%%%%%
\begin{figure}[tbp]
\includegraphics[width=9 cm]{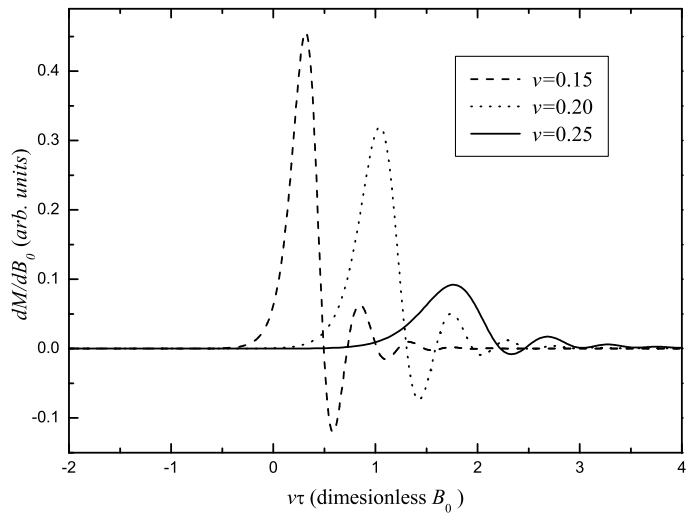}
\caption{$dM/dB_{0}$ as a function of the dimensionless external magnetic
field for different sweep rates. $\protect\gamma =0.1,$ $\protect\kappa =1$
and the dynamics is governed by Eqs.~(\protect\ref{hp},\protect\ref{Zp},%
\protect\ref{Rp}).}
\label{bswfig}
\end{figure}
%%%%%%%%%%%%%%%%%%%%%%%%%%%%%%%%%%%%%%%%%%%%%%%%%%%%%%%%%%%%%%%%%%%%%%%%%%%
An important experimental result is that the position of the peaks in
$dM/dB_{0}$ corresponding to the radiation process does depend on the external
field sweep rate $v$. In our model this is related to the time spent by the
sample around the resonance. For a slowly changing $B_{0}$, the dynamics is
similar to the case of constant detuning, where an analytical solution is
known, while increasing the value of $v$, an appropriate numerical solution of
the dynamics is needed. In the case of superradiance, Fig.~\ref{bswfig} shows
$dM/dB_{0}$ as a function of $v\tau $, i.e., the dimensionless
external field $B_{0}$. As we can see, for larger values of the sweep rate $%
v $, the height of the emission peak decreases and its position is shifted
towards higher field values. The shift is in agreement with the experimental
findings, while as a consequence of the coherent interaction, $dM/dB_{0}$
exhibits oscillations with sign changes.

%%%%%%%%%%%%%%%%%%%%%%%%%%%%%%%%%%%%%%%%%%%%%%%%%%%%%%%%%%%%%%%%%%%%%%%%%%%
\begin{figure}[tbp]
\includegraphics[width=9 cm]{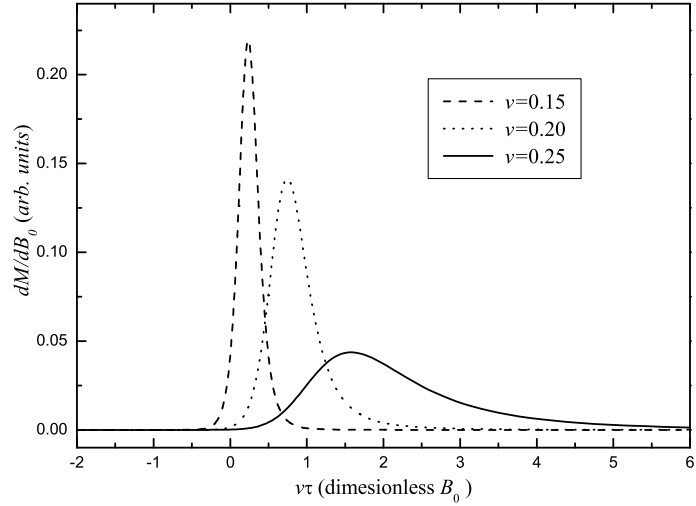}
\caption{$dM/dB_0$ as a function of the dimensionless external magnetic field
for different sweep rates. $\protect\gamma =1,$ $\protect\kappa =0.1$ but in
contrast with Fig.~\protect\ref{bswfig}, the rate equations
(\protect\ref{rate}) have been used to calculate the time evolution.}
\label{bratefig}
\end{figure}
%%%%%%%%%%%%%%%%%%%%%%%%%%%%%%%%%%%%%%%%%%%%%%%%%%%%%%%%%%%%%%%%%%%%%%%%%%%

If we assume that the maser effect is responsible for the radiation and use
the rate equations (\ref{rate}) to calculate the dynamics of the system,
somewhat different results are obtained. As Fig.~\ref{bratefig} shows,
$dM/dB_{0}$ accompanied with maser radiation, scales similarly with $v$ as in
the case of weak dephasing: larger sweep rates correspond to peaks at higher
$B_{0}$ fields, thus taking the cavity effects into account, this scaling
property is not characteristic for SR. However, the oscillations seen in the
superradiant case are absent in Fig.~\ref{bratefig}.
\bigskip

We shall analyze now from the point of view of transversal relaxation if the
observed radiation could be superradiance. The reduction procedure summarized
in Sec.~\ref{levelsection} allows us to calculate the matrix element $|s|$ and
thus the characteristic time (\ref{T0def}) of the emission process for any
transition. As the experimentally observed radiation peaks were around $1.4$
T, we focus on this value of the external magnetic field. The level structure
of the Hamiltonian (\ref{H}) provides the resonant transition frequency for a
given transition $m\rightarrow m^{\prime }$, as well as the population of the
upper level according to the Boltzmann factor. In this way we can calculate
$T_{0}$ for any transition as a function of the temperature of the sample. As
for SR to occur the dephasing rate $\gamma =T_{0}/T_{2}$ must be small, so we
should look for the transition with the minimal value of $T_{0}$. We found
that below approx.~$0.8$ K the transition $m=-10\rightarrow m^{\prime }=8$
provides the shortest $T_{0}$, while above this temperature the transition
from $m=-6$ to $m^{\prime }=4$ yields the minimal characteristic time. As
Fig.~\ref{T0fig}
shows, for low temperatures, i.e., ground state tunneling, the minimal $%
T_{0} $ is of the order of seconds. This is a consequence of the very small
coupling coefficient $s$, corresponding to this transition. For higher
temperatures when the transition $m=-6\rightarrow m^{\prime }=4$ provides the
shortest characteristic time, $T_{0}$ significantly decreases as a function of
temperature. This is a consequence of the relation $T_{0}\propto
1/(|s|\sqrt{N_{0}})$  (see Eq.~(\ref{T0def})), where $N_{0}$ is temperature
dependent. That is, above $0.8$ K, the most favorable conditions for
superradiance might be realized in the case of the transition $m=-6\rightarrow
m^{\prime }=4,$ with a still strong temperature dependence of $T_{0}$.

However, the energy emitted during a transition process is not necessarily the
highest for the lowest $T_{0}$. In fact, the population of the $m=-6$ level --
which determines the maximum number of the active molecules -- is not large
enough to explain the magnitude of the emitted energy observed in a recent
experiment. Ref.~[{\onlinecite{HJAGHT05}] reports on radiative bursts of
duration of a few milliseconds, where (at 2 K) the total energy emitted by the
sample was detected to be around $3$ \emph{nJ}. Using the parameters of the
experiment,\cite{HJAGHT05} we investigated all the possible transitions and
found the best agreement with the experimental data for the transition
$m=-8\rightarrow m^{\prime }=6$, giving a value of $T_{0}$ in the \emph{ms}
range and a total emitted energy to be around 1.5 \emph{nJ}. (Note that for
initial states below $m=-8$ the time scale of the process turns out to be too
long, while for $m>-8$  the number of active molecules is too small.) Thus our
model predicts that the process having the most important role in producing
the observed radiation is the transition $m=-8\rightarrow m^{\prime }=6.$

%%%%%%%%%%%%%%%%%%%%%%%%%%%%%%%%%%%%%%%%%%%%%%%%%%%%%%%%%%%%%%%%%%%%%%%%%%%
\begin{figure}[tbp]
\includegraphics[width=9 cm]{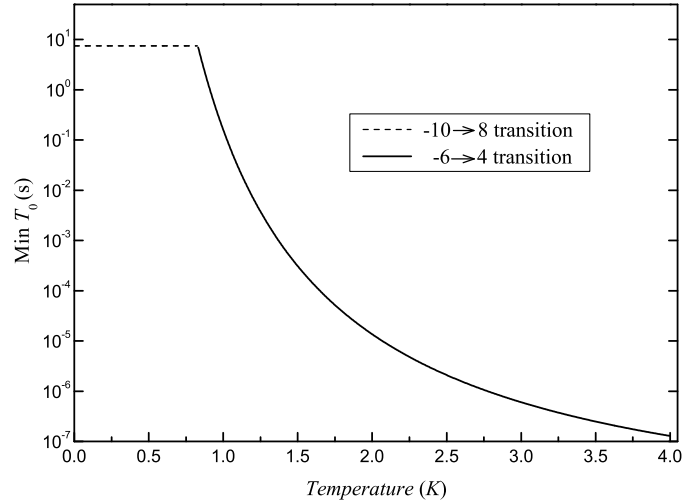}
\caption{The minimal value (over the possible transitions) of $T_0$ given by
Eq.~(\protect\ref{T0def}) at $B_0=1.4$ T, where microwave emission from the
sample has been detected. The minimum at low temperatures corresponds to the
transition $m=-10 \rightarrow m^{\prime }=8$, while above approx.~$0.8$ K $%
m=-6 \rightarrow m^{\prime }=4$ provides the shortest $T_0.$}
\label{T0fig}
\end{figure}
%%%%%%%%%%%%%%%%%%%%%%%%%%%%%%%%%%%%%%%%%%%%%%%%%%%%%%%%%%%%%%%%%%%%%%%%%%%

As we have seen, the character of the emission depends on the ratio $\gamma
=T_{0}/T_{2},$ where $T_{2}$ is decreasing with increasing temperature.
According to Fig.~\ref{T0fig}, even for the shortest possible $T_{0}$ and
relatively weak dephasing, with $T_{2}$ around $10^{-5}$--$10^{-6}$ \emph{s}
(Ref.~[{\onlinecite{LL99}]), we have $\gamma >1$ at $2$ K. The millisecond
time scale obtained here and observed also in the experiments is clearly
longer than the relaxation time $T_{2}$. Thus -- unless a yet unknown effect
decreases the disturbance caused by phase relaxation -- the process
responsible for the experimentally observed\cite{TCHA04} bursts of radiation
seems to be rather a maser effect than superradiance.

The fact that no emission was seen for external fields lower than 1.4 T (where
there can be a resonance as well) can be explained by the strength of the
coupling to the transversal mode: $|s|$ is generally at least an order of
magnitude larger for the possibly relevant transitions at $1.4$ T than at
previous resonances. However, in a good resonator -- like the Fabry-Perot
mirrors in Ref.~[{\onlinecite{TAH02}] -- one may expect that several
resonances have observable consequences, radiation and -- as it has been
detected -- enhanced magnetic relaxation rates. Additionally, we note that at
high sweep rates\cite{VSM04} the system passes not only a single cavity
resonance during the emission process, and consecutive resonances can broaden
the peaks in the $dM/dB_{0}(B_{0})$ plots.
%%%%%%%%%%%%%%%%%%%%%%%%%%%%%%%%%%%%%%%%%%%%%%%%%%%%%%%%%%%%%%%%%%%%%%%%%%%

\section{Summary and Conclusions}

\label{summarysection} In this paper we developed a model for the interaction
of a crystal of molecular magnets with the magnetic field of a surrounding
cavity. The sample itself generates this transversal field $B$, while it also
acts back on the molecules. The most important point of our treatment is that
the cavity mode with fixed frequency $\Omega $ comes to resonance with a
magnetic transition at a given value of the external longitudinal magnetic
field. Around this resonance the interaction of the molecules with the mode
significantly increases leading to an observable burst of electromagnetic
radiation as well as a change in the magnetization of the sample. Our model
can describe different mechanisms of this radiation, in fact, there is a
continuous transition from superradiance to maser-like effects. The crucial
parameter here is the ratio of two time
scales, the characteristic time of the process and the dephasing time $%
\gamma =T_{0}/T_{2}$. For small values of $\gamma $ the time evolution of the
molecules is coherent allowing for the strong collective effect of
superradiance in a cavity. In the case of strong dephasing, the sample still
can emit electromagnetic radiation, but now the coherence of the molecules
plays no role, the maser rate equations with time dependent detuning can
describe the process. For moderate values of $\gamma $ we have a transition
between the two processes. By calculating the intensity of the emitted
radiation we have shown that with increasing the sweep rate of the external
magnetic field, the emission peaks are shifted towards higher field values in
accordance with the experimental results. This statement holds for both
emission mechanisms, but the detailed functional dependencies are different
for SR and maser emission.

Based on realistic approximations for $T_{0}$ and $T_{2}$, the process
responsible for the experimentally observed bursts of electromagnetic
radiation is most probably not superradiance, but rather a maser effect. The
comparison of time resolved experiments on the emitted radiation with our
theoretical results would provide the necessary information in order to settle
this question. We expect that at very low temperatures, when spin-phonon
relaxation is weaker, the collective features of the radiation may become
dominant. While this is an interesting problem on its own, it is expected that
the analysis of the radiated field can yield additional information on the
process of quantum tunneling, as well as on the detailed properties of the
interaction of these crystals and the field.

\acknowledgments We thank O. K\'{a}lm\'{a}n for her valuable comments. This
work was supported by the Flemish-Hungarian Bilateral Programme, the Flemish
Science Foundation (FWO-Vl), the Belgian Science Policy and the Hungarian
Scientific Research Fund (OTKA) under Contracts Nos. T48888, D46043, M36803,
M045596.


\begin{thebibliography}{43}
\expandafter\ifx\csname natexlab\endcsname\relax\def\natexlab#1{#1}\fi
\expandafter\ifx\csname bibnamefont\endcsname\relax
  \def\bibnamefont#1{#1}\fi
\expandafter\ifx\csname bibfnamefont\endcsname\relax
  \def\bibfnamefont#1{#1}\fi
\expandafter\ifx\csname citenamefont\endcsname\relax
  \def\citenamefont#1{#1}\fi
\expandafter\ifx\csname url\endcsname\relax
  \def\url#1{\texttt{#1}}\fi
\expandafter\ifx\csname urlprefix\endcsname\relax\def\urlprefix{URL }\fi
\providecommand{\bibinfo}[2]{#2} \providecommand{\eprint}[2][]{\url{#2}}

\bibitem[{\citenamefont{Barbara et~al.}(1999)\citenamefont{Barbara, Thomas,
  Lionti, Chiorescu, and Sulpice}}]{BTLCS99}
\bibinfo{author}{\bibfnamefont{B.}~\bibnamefont{Barbara}},
  \bibinfo{author}{\bibfnamefont{L.}~\bibnamefont{Thomas}},
  \bibinfo{author}{\bibfnamefont{F.}~\bibnamefont{Lionti}},
  \bibinfo{author}{\bibfnamefont{I.}~\bibnamefont{Chiorescu}},
  \bibnamefont{and} \bibinfo{author}{\bibfnamefont{A.}~\bibnamefont{Sulpice}},
  \bibinfo{journal}{J. Magn. Magn. Mat.} \textbf{\bibinfo{volume}{200}},
  \bibinfo{pages}{167} (\bibinfo{year}{1999}).

\bibitem[{\citenamefont{Wernsdorfer and Sessoli}(1999)}]{WS99}
\bibinfo{author}{\bibfnamefont{W.}~\bibnamefont{Wernsdorfer}} \bibnamefont{and}
  \bibinfo{author}{\bibfnamefont{R.}~\bibnamefont{Sessoli}},
  \bibinfo{journal}{Science} \textbf{\bibinfo{volume}{284}},
  \bibinfo{pages}{133} (\bibinfo{year}{1999}).

\bibitem[{\citenamefont{Villain}(2003)}]{V03}
\bibinfo{author}{\bibfnamefont{J.}~\bibnamefont{Villain}},
  \bibinfo{journal}{Annales de Physique} \textbf{\bibinfo{volume}{28}},
  \bibinfo{pages}{1} (\bibinfo{year}{2003}).

\bibitem[{\citenamefont{Friedman et~al.}(1996)\citenamefont{Friedman, Sarachik,
  Tejada, and Ziolo}}]{FST96}
\bibinfo{author}{\bibfnamefont{J.~R.} \bibnamefont{Friedman}},
  \bibinfo{author}{\bibfnamefont{M.~P.} \bibnamefont{Sarachik}},
  \bibinfo{author}{\bibfnamefont{J.}~\bibnamefont{Tejada}}, \bibnamefont{and}
  \bibinfo{author}{\bibfnamefont{R.}~\bibnamefont{Ziolo}},
  \bibinfo{journal}{Phys. Rev. Lett.} \textbf{\bibinfo{volume}{76}},
  \bibinfo{pages}{3830} (\bibinfo{year}{1996}).

\bibitem[{\citenamefont{Chudnovsky and Garanin}(2002)}]{CG02}
\bibinfo{author}{\bibfnamefont{E.~M.} \bibnamefont{Chudnovsky}}
  \bibnamefont{and} \bibinfo{author}{\bibfnamefont{D.~A.}
  \bibnamefont{Garanin}}, \bibinfo{journal}{Phys. Rev. Lett.}
  \textbf{\bibinfo{volume}{89}}, \bibinfo{pages}{157201}
  (\bibinfo{year}{2002}).

\bibitem[{\citenamefont{Henner and Kaganov}(2003)}]{HK03}
\bibinfo{author}{\bibfnamefont{V.~K.} \bibnamefont{Henner}} \bibnamefont{and}
  \bibinfo{author}{\bibfnamefont{I.~V.} \bibnamefont{Kaganov}},
  \bibinfo{journal}{Phys. Rev. B} \textbf{\bibinfo{volume}{68}},
  \bibinfo{pages}{144420} (\bibinfo{year}{2003}).

\bibitem[{\citenamefont{Tejada et~al.}(2004)\citenamefont{Tejada, Chudnovsky,
  Hernandez, and Amig\'{o}}}]{TCHA04}
\bibinfo{author}{\bibfnamefont{J.}~\bibnamefont{Tejada}},
  \bibinfo{author}{\bibfnamefont{E.~M.} \bibnamefont{Chudnovsky}},
  \bibinfo{author}{\bibfnamefont{J.~M.} \bibnamefont{Hernandez}},
  \bibnamefont{and}
  \bibinfo{author}{\bibfnamefont{R.}~\bibnamefont{Amig\'{o}}},
  \bibinfo{journal}{Appl. Phys. Lett.} \textbf{\bibinfo{volume}{84}},
  \bibinfo{pages}{2373} (\bibinfo{year}{2004}).

\bibitem[{\citenamefont{Vanacken et~al.}(2004)\citenamefont{Vanacken,
  Stroobants, Malfait, Moschalkov, Jordi, Tejada, Amigo, Chudnovsky, and
  Garanin}}]{VSM04}
\bibinfo{author}{\bibfnamefont{J.}~\bibnamefont{Vanacken}},
  \bibinfo{author}{\bibfnamefont{S.}~\bibnamefont{Stroobants}},
  \bibinfo{author}{\bibfnamefont{M.}~\bibnamefont{Malfait}},
  \bibinfo{author}{\bibfnamefont{V.~V.} \bibnamefont{Moschalkov}},
  \bibinfo{author}{\bibfnamefont{M.}~\bibnamefont{Jordi}},
  \bibinfo{author}{\bibfnamefont{J.}~\bibnamefont{Tejada}},
  \bibinfo{author}{\bibfnamefont{R.}~\bibnamefont{Amigo}},
  \bibinfo{author}{\bibfnamefont{E.~M.} \bibnamefont{Chudnovsky}},
  \bibnamefont{and} \bibinfo{author}{\bibfnamefont{D.~A.}
  \bibnamefont{Garanin}}, \bibinfo{journal}{Phys. Rev. B}
  \textbf{\bibinfo{volume}{70}}, \bibinfo{pages}{220401}
  (\bibinfo{year}{2004}).

\bibitem[{\citenamefont{Hernandez-Minguez
  et~al.}(2005)\citenamefont{Hernandez-Minguez, Jordi, Amigo, Garcia-Santiago,
  Hernandez, and Tejada}}]{HJAGHT05}
\bibinfo{author}{\bibfnamefont{A.}~\bibnamefont{Hernandez-Minguez}},
  \bibinfo{author}{\bibfnamefont{M.}~\bibnamefont{Jordi}},
  \bibinfo{author}{\bibfnamefont{R.}~\bibnamefont{Amigo}},
  \bibinfo{author}{\bibfnamefont{A.}~\bibnamefont{Garcia-Santiago}},
  \bibinfo{author}{\bibfnamefont{J.~M.} \bibnamefont{Hernandez}},
  \bibnamefont{and} \bibinfo{author}{\bibfnamefont{J.}~\bibnamefont{Tejada}},
  \bibinfo{journal}{Europhys. Lett.} \textbf{\bibinfo{volume}{69}},
  \bibinfo{pages}{270} (\bibinfo{year}{2005}).

\bibitem[{\citenamefont{Joseph et~al.}(2004)\citenamefont{Joseph, Calero, and
  Chudnovsky}}]{JCC04}
\bibinfo{author}{\bibfnamefont{C.~L.} \bibnamefont{Joseph}},
  \bibinfo{author}{\bibfnamefont{C.}~\bibnamefont{Calero}}, \bibnamefont{and}
  \bibinfo{author}{\bibfnamefont{E.~M.} \bibnamefont{Chudnovsky}}
  (\bibinfo{year}{2004}).

\bibitem[{\citenamefont{Dicke}(1954)}]{D54}
\bibinfo{author}{\bibfnamefont{R.~M.} \bibnamefont{Dicke}},
  \bibinfo{journal}{Phys. Rev.} \textbf{\bibinfo{volume}{93}},
  \bibinfo{pages}{439} (\bibinfo{year}{1954}).

\bibitem[{\citenamefont{Benedict et~al.}(1996)\citenamefont{Benedict, Ermolaev,
  Malyshev, Sokolov, and Trifonov}}]{SRAD}
\bibinfo{author}{\bibfnamefont{M.~G.} \bibnamefont{Benedict}},
  \bibinfo{author}{\bibfnamefont{A.~M.} \bibnamefont{Ermolaev}},
  \bibinfo{author}{\bibfnamefont{V.~A.} \bibnamefont{Malyshev}},
  \bibinfo{author}{\bibfnamefont{I.~V.} \bibnamefont{Sokolov}},
  \bibnamefont{and} \bibinfo{author}{\bibfnamefont{E.~D.}
  \bibnamefont{Trifonov}}, \emph{\bibinfo{title}{Superradiance}}
  (\bibinfo{publisher}{IOP}, \bibinfo{address}{Bristol}, \bibinfo{year}{1996}).

\bibitem[{\citenamefont{Gross and Haroche}(1982)}]{GH82}
\bibinfo{author}{\bibfnamefont{M.}~\bibnamefont{Gross}} \bibnamefont{and}
  \bibinfo{author}{\bibfnamefont{S.}~\bibnamefont{Haroche}},
  \bibinfo{journal}{Phys. Rep.} \textbf{\bibinfo{volume}{93}},
  \bibinfo{pages}{301 } (\bibinfo{year}{1982}).

\bibitem[{\citenamefont{Kiselev et~al.}(1988)\citenamefont{Kiselev,
  Prudkoglyad, Shumovsky, and Yukalov}}]{KPSY88}
\bibinfo{author}{\bibfnamefont{Y.~F.} \bibnamefont{Kiselev}},
  \bibinfo{author}{\bibfnamefont{A.~F.} \bibnamefont{Prudkoglyad}},
  \bibinfo{author}{\bibfnamefont{A.~S.} \bibnamefont{Shumovsky}},
  \bibnamefont{and} \bibinfo{author}{\bibfnamefont{V.~I.}
  \bibnamefont{Yukalov}}, \bibinfo{journal}{Sov. Phys.: JETP}
  \textbf{\bibinfo{volume}{67}}, \bibinfo{pages}{413} (\bibinfo{year}{1988}).

\bibitem[{\citenamefont{Bazhanov et~al.}(1990)\citenamefont{Bazhanov,
  Bulyanitsa, Zaitsev, Kovalev, Malyshev, and Trifonov}}]{BBZKMT90}
\bibinfo{author}{\bibfnamefont{N.~A.} \bibnamefont{Bazhanov}},
  \bibinfo{author}{\bibfnamefont{D.~S.} \bibnamefont{Bulyanitsa}},
  \bibinfo{author}{\bibfnamefont{A.~I.} \bibnamefont{Zaitsev}},
  \bibinfo{author}{\bibfnamefont{A.~I.} \bibnamefont{Kovalev}},
  \bibinfo{author}{\bibfnamefont{V.~A.} \bibnamefont{Malyshev}},
  \bibnamefont{and} \bibinfo{author}{\bibfnamefont{E.~D.}
  \bibnamefont{Trifonov}}, \bibinfo{journal}{Sov. Phys.: JETP}
  \textbf{\bibinfo{volume}{70}}, \bibinfo{pages}{1128} (\bibinfo{year}{1990}).

\bibitem[{\citenamefont{Yukalov and Yukalova}(2004)}]{YY04}
\bibinfo{author}{\bibfnamefont{V.~I.} \bibnamefont{Yukalov}} \bibnamefont{and}
  \bibinfo{author}{\bibfnamefont{E.~P.} \bibnamefont{Yukalova}},
  \bibinfo{journal}{Phys. Part. Nucl.} \textbf{\bibinfo{volume}{35}},
  \bibinfo{pages}{348} (\bibinfo{year}{2004}).

\bibitem[{\citenamefont{Haroche and Raimond}(1985)}]{HR85}
\bibinfo{author}{\bibfnamefont{S.}~\bibnamefont{Haroche}} \bibnamefont{and}
  \bibinfo{author}{\bibfnamefont{J.~M.} \bibnamefont{Raimond}}
  (\bibinfo{year}{1985}), vol.~\bibinfo{volume}{20} of
  \emph{\bibinfo{series}{Advances in atomic and molecular physics}}, pp.
  \bibinfo{pages}{347--411}.

\bibitem[{\citenamefont{Bloembergen and Pound}(1954)}]{BP54}
\bibinfo{author}{\bibfnamefont{N.}~\bibnamefont{Bloembergen}} \bibnamefont{and}
  \bibinfo{author}{\bibfnamefont{R.~V.} \bibnamefont{Pound}},
  \bibinfo{journal}{Phys. Rev.} \textbf{\bibinfo{volume}{95}},
  \bibinfo{pages}{8} (\bibinfo{year}{1954}).

\bibitem[{\citenamefont{Yukalov and Yukalova}(2005{\natexlab{a}})}]{YY05a}
\bibinfo{author}{\bibfnamefont{V.~I.} \bibnamefont{Yukalov}} \bibnamefont{and}
  \bibinfo{author}{\bibfnamefont{E.~P.} \bibnamefont{Yukalova}},
  \bibinfo{journal}{Laser Phys. Lett.} \textbf{\bibinfo{volume}{2}},
  \bibinfo{pages}{302} (\bibinfo{year}{2005}{\natexlab{a}}).

\bibitem[{\citenamefont{Yukalov and Yukalova}(2005{\natexlab{b}})}]{YY05b}
\bibinfo{author}{\bibfnamefont{V.~I.} \bibnamefont{Yukalov}} \bibnamefont{and}
  \bibinfo{author}{\bibfnamefont{E.~P.} \bibnamefont{Yukalova}},
  \bibinfo{journal}{Europhys. Lett.} \textbf{\bibinfo{volume}{70}},
  \bibinfo{pages}{306} (\bibinfo{year}{2005}{\natexlab{b}}).

\bibitem[{\citenamefont{Tejada et~al.}(2003)\citenamefont{Tejada, Amigo,
  Hernandez, and Chudnovsky}}]{TAH02}
\bibinfo{author}{\bibfnamefont{J.}~\bibnamefont{Tejada}},
  \bibinfo{author}{\bibfnamefont{R.}~\bibnamefont{Amigo}},
  \bibinfo{author}{\bibfnamefont{J.~M.} \bibnamefont{Hernandez}},
  \bibnamefont{and} \bibinfo{author}{\bibfnamefont{E.~M.}
  \bibnamefont{Chudnovsky}}, \bibinfo{journal}{Phys. Rev. B}
  \textbf{\bibinfo{volume}{68}}, \bibinfo{pages}{014431}
  (\bibinfo{year}{2003}).

\bibitem[{\citenamefont{Leuenberger and Loss}(2001)}]{LL01}
\bibinfo{author}{\bibfnamefont{M.~N.} \bibnamefont{Leuenberger}}
  \bibnamefont{and} \bibinfo{author}{\bibfnamefont{D.}~\bibnamefont{Loss}},
  \bibinfo{journal}{Nature} \textbf{\bibinfo{volume}{410}},
  \bibinfo{pages}{789} (\bibinfo{year}{2001}).

\bibitem[{\citenamefont{Mertes et~al.}(2001)\citenamefont{Mertes, Suzuki,
  Sarachik, Paltiel, Shtrikman, Zeldov, Rumberger, Hendrickson, and
  Christou}}]{MSS01}
\bibinfo{author}{\bibfnamefont{K.~M.} \bibnamefont{Mertes}},
  \bibinfo{author}{\bibfnamefont{Y.}~\bibnamefont{Suzuki}},
  \bibinfo{author}{\bibfnamefont{M.~P.} \bibnamefont{Sarachik}},
  \bibinfo{author}{\bibfnamefont{Y.}~\bibnamefont{Paltiel}},
  \bibinfo{author}{\bibfnamefont{H.}~\bibnamefont{Shtrikman}},
  \bibinfo{author}{\bibfnamefont{E.}~\bibnamefont{Zeldov}},
  \bibinfo{author}{\bibfnamefont{E.}~\bibnamefont{Rumberger}},
  \bibinfo{author}{\bibfnamefont{D.~N.} \bibnamefont{Hendrickson}},
  \bibnamefont{and} \bibinfo{author}{\bibfnamefont{G.}~\bibnamefont{Christou}},
  \bibinfo{journal}{Phys. Rev. Lett.} \textbf{\bibinfo{volume}{87}},
  \bibinfo{pages}{227205} (\bibinfo{year}{2001}).

\bibitem[{\citenamefont{Mirebeau et~al.}(1999)\citenamefont{Mirebeau, Hennion,
  Casalta, Andres, G\"{u}del, Irodova, and Caneschi}}]{M99}
\bibinfo{author}{\bibfnamefont{I.}~\bibnamefont{Mirebeau}},
  \bibinfo{author}{\bibfnamefont{M.}~\bibnamefont{Hennion}},
  \bibinfo{author}{\bibfnamefont{H.}~\bibnamefont{Casalta}},
  \bibinfo{author}{\bibfnamefont{H.}~\bibnamefont{Andres}},
  \bibinfo{author}{\bibfnamefont{H.~U.} \bibnamefont{G\"{u}del}},
  \bibinfo{author}{\bibfnamefont{A.~V.} \bibnamefont{Irodova}},
  \bibnamefont{and} \bibinfo{author}{\bibfnamefont{A.}~\bibnamefont{Caneschi}},
  \bibinfo{journal}{Phys. Rev. Lett.} \textbf{\bibinfo{volume}{83}},
  \bibinfo{pages}{628} (\bibinfo{year}{1999}).

\bibitem[{\citenamefont{Barra et~al.}(1997)\citenamefont{Barra, Gatteschi, and
  Sessoli}}]{BGS97}
\bibinfo{author}{\bibfnamefont{A.~L.} \bibnamefont{Barra}},
  \bibinfo{author}{\bibfnamefont{D.}~\bibnamefont{Gatteschi}},
  \bibnamefont{and} \bibinfo{author}{\bibfnamefont{R.}~\bibnamefont{Sessoli}},
  \bibinfo{journal}{Phys. Rev. B} \textbf{\bibinfo{volume}{56}},
  \bibinfo{pages}{8192} (\bibinfo{year}{1997}).

\bibitem[{\citenamefont{Hill et~al.}(1998)\citenamefont{Hill, Perenboom, Dalal,
  Hathaway, Stalcup, and Brooks}}]{H98}
\bibinfo{author}{\bibfnamefont{S.}~\bibnamefont{Hill}},
  \bibinfo{author}{\bibfnamefont{J.~A. A.~J.} \bibnamefont{Perenboom}},
  \bibinfo{author}{\bibfnamefont{N.~S.} \bibnamefont{Dalal}},
  \bibinfo{author}{\bibfnamefont{T.}~\bibnamefont{Hathaway}},
  \bibinfo{author}{\bibfnamefont{T.}~\bibnamefont{Stalcup}}, \bibnamefont{and}
  \bibinfo{author}{\bibfnamefont{J.~S.} \bibnamefont{Brooks}},
  \bibinfo{journal}{Phys. Rev. Lett.} \textbf{\bibinfo{volume}{80}},
  \bibinfo{pages}{2453} (\bibinfo{year}{1998}).

\bibitem[{\citenamefont{Hill et~al.}(2003)\citenamefont{Hill, Edwards, Jones,
  Dalal, and North}}]{HEJ03}
\bibinfo{author}{\bibfnamefont{S.}~\bibnamefont{Hill}},
  \bibinfo{author}{\bibfnamefont{R.~S.} \bibnamefont{Edwards}},
  \bibinfo{author}{\bibfnamefont{S.~I.} \bibnamefont{Jones}},
  \bibinfo{author}{\bibfnamefont{N.~S.} \bibnamefont{Dalal}}, \bibnamefont{and}
  \bibinfo{author}{\bibfnamefont{J.~M.} \bibnamefont{North}},
  \bibinfo{journal}{Phys. Rev. Lett.} \textbf{\bibinfo{volume}{90}},
  \bibinfo{pages}{217204} (\bibinfo{year}{2003}).

\bibitem[{\citenamefont{van Vleck}(1929)}]{vV29}
\bibinfo{author}{\bibfnamefont{J.~H.} \bibnamefont{van Vleck}},
  \bibinfo{journal}{Phys. Rev.} \textbf{\bibinfo{volume}{33}},
  \bibinfo{pages}{467} (\bibinfo{year}{1929}).

\bibitem[{\citenamefont{des Cloizeaux}(1960)}]{dC60}
\bibinfo{author}{\bibfnamefont{J.}~\bibnamefont{des Cloizeaux}},
  \bibinfo{journal}{Nucl. Phys.} \textbf{\bibinfo{volume}{20}},
  \bibinfo{pages}{321} (\bibinfo{year}{1960}).

\bibitem[{\citenamefont{Klein}(1974)}]{K74}
\bibinfo{author}{\bibfnamefont{D.~J.} \bibnamefont{Klein}},
  \bibinfo{journal}{J. Chem. Phys} \textbf{\bibinfo{volume}{61}},
  \bibinfo{pages}{786} (\bibinfo{year}{1974}).

\bibitem[{\citenamefont{Garanin}(1991)}]{G91}
\bibinfo{author}{\bibfnamefont{D.~A.} \bibnamefont{Garanin}},
  \bibinfo{journal}{J. Phys. A} \textbf{\bibinfo{volume}{24}},
  \bibinfo{pages}{L61} (\bibinfo{year}{1991}).

\bibitem[{\citenamefont{Leuenberger and Loss}(2000)}]{LL00}
\bibinfo{author}{\bibfnamefont{M.~N.} \bibnamefont{Leuenberger}}
  \bibnamefont{and} \bibinfo{author}{\bibfnamefont{D.}~\bibnamefont{Loss}},
  \bibinfo{journal}{Phys. Rev. B} \textbf{\bibinfo{volume}{61}},
  \bibinfo{pages}{1286} (\bibinfo{year}{2000}).

\bibitem[{\citenamefont{Yoo and Park}(2005)}]{YP05}
\bibinfo{author}{\bibfnamefont{S.-K.} \bibnamefont{Yoo}} \bibnamefont{and}
  \bibinfo{author}{\bibfnamefont{C.-S.} \bibnamefont{Park}},
  \bibinfo{journal}{Phys. Rev. B} \textbf{\bibinfo{volume}{71}},
  \bibinfo{pages}{012409} (\bibinfo{year}{2005}).

\bibitem[{\citenamefont{Landau}(1932)}]{L32}
\bibinfo{author}{\bibfnamefont{L.~D.} \bibnamefont{Landau}},
  \bibinfo{journal}{Phys. Z. Sowjetunion} \textbf{\bibinfo{volume}{2}},
  \bibinfo{pages}{46} (\bibinfo{year}{1932}).

\bibitem[{\citenamefont{Zener}(1932)}]{Z32}
\bibinfo{author}{\bibfnamefont{C.}~\bibnamefont{Zener}},
  \bibinfo{journal}{Proc. Roy. Soc. London, Ser. A}
  \textbf{\bibinfo{volume}{137}}, \bibinfo{pages}{696} (\bibinfo{year}{1932}).

\bibitem[{\citenamefont{St\"{u}ckelberg}(1932)}]{S32}
\bibinfo{author}{\bibfnamefont{E.~C.~G.} \bibnamefont{St\"{u}ckelberg}},
  \bibinfo{journal}{Helv. Phys. Acta} \textbf{\bibinfo{volume}{5}},
  \bibinfo{pages}{369} (\bibinfo{year}{1932}).

\bibitem[{\citenamefont{Sargent et~al.}(1974)\citenamefont{Sargent, Scully, and
  Lamb}}]{SSL74}
\bibinfo{author}{\bibfnamefont{M.}~\bibnamefont{Sargent}},
  \bibinfo{author}{\bibfnamefont{M.~O.} \bibnamefont{Scully}},
  \bibnamefont{and} \bibinfo{author}{\bibfnamefont{W.~E.} \bibnamefont{Lamb}},
  \emph{\bibinfo{title}{Laser Physics}} (\bibinfo{publisher}{Reading MA:
  Addison-Wesley}, \bibinfo{year}{1974}).

\bibitem[{\citenamefont{Jackson}(1998)}]{J98}
\bibinfo{author}{\bibfnamefont{J.~D.} \bibnamefont{Jackson}},
  \emph{\bibinfo{title}{Classical Electrodynamics}} (\bibinfo{publisher}{John
  Wiley \& Sons}, \bibinfo{year}{1998}), \bibinfo{edition}{3rd} ed.

\bibitem[{\citenamefont{Friedberg et~al.}(1973)\citenamefont{Friedberg,
  Hartmann, and Manassah}}]{FHM73}
\bibinfo{author}{\bibfnamefont{R.}~\bibnamefont{Friedberg}},
  \bibinfo{author}{\bibfnamefont{S.~R.} \bibnamefont{Hartmann}},
  \bibnamefont{and} \bibinfo{author}{\bibfnamefont{J.~T.}
  \bibnamefont{Manassah}}, \bibinfo{journal}{Phys. Rep. C}
  \textbf{\bibinfo{volume}{7}}, \bibinfo{pages}{101} (\bibinfo{year}{1973}).

\bibitem[{\citenamefont{Zaitsev et~al.}(1983)\citenamefont{Zaitsev, Malyshev,
  and Trifonov}}]{ZMT83}
\bibinfo{author}{\bibfnamefont{A.~I.} \bibnamefont{Zaitsev}},
  \bibinfo{author}{\bibfnamefont{V.~A.} \bibnamefont{Malyshev}},
  \bibnamefont{and} \bibinfo{author}{\bibfnamefont{E.~D.}
  \bibnamefont{Trifonov}}, \bibinfo{journal}{Sov. Phys.: JETP}
  \textbf{\bibinfo{volume}{57}}, \bibinfo{pages}{275} (\bibinfo{year}{1983}).

\bibitem[{\citenamefont{Davis et~al.}(2005)\citenamefont{Davis, Henner,
  Tchernatinsky, and Kaganov}}]{DHTK05}
\bibinfo{author}{\bibfnamefont{C.~L.} \bibnamefont{Davis}},
  \bibinfo{author}{\bibfnamefont{V.~K.} \bibnamefont{Henner}},
  \bibinfo{author}{\bibfnamefont{A.~V.} \bibnamefont{Tchernatinsky}},
  \bibnamefont{and} \bibinfo{author}{\bibfnamefont{I.~V.}
  \bibnamefont{Kaganov}}, \bibinfo{journal}{Phys. Rev. B}
  \textbf{\bibinfo{volume}{72}}, \bibinfo{pages}{054406}
  (\bibinfo{year}{2005}).

\bibitem[{\citenamefont{Leuenberger and Loss}(1999)}]{LL99}
\bibinfo{author}{\bibfnamefont{M.~N.} \bibnamefont{Leuenberger}}
  \bibnamefont{and} \bibinfo{author}{\bibfnamefont{D.}~\bibnamefont{Loss}},
  \bibinfo{journal}{Europhys. Lett.} \textbf{\bibinfo{volume}{45}},
  \bibinfo{pages}{692} (\bibinfo{year}{1999}).

\bibitem[{\citenamefont{Sessoli}(1995)}]{Se95}
\bibinfo{author}{\bibfnamefont{R.}~\bibnamefont{Sessoli}},
  \bibinfo{journal}{Mol. Cryst. Liq. Cryst. Sci. Technol., Sect A}
  \textbf{\bibinfo{volume}{274}}, \bibinfo{pages}{145} (\bibinfo{year}{1995}).

\end{thebibliography}
\end{document}